\documentclass[11pt]{article}
\begin{document}
\def\p{\partial}
\def\f{\frac}
\def\l{\left}
\def\r{\right}
\def\({\l(}
\def\){\r)}

\baselineskip 22pt
\begin{center}
{\Large \bf Quantum Left-handed metamaterial with superconducting
quantum interference devices } \vspace{1.0cm}

\renewcommand{\thefootnote}{$\dag$}

 Chunguang Du\footnote{{\bf
E-mail}: duchunguang@tsinghua.org.cn} \hspace{2cm} Hongyi Chen
\hspace{2cm} Shiqun Li \vspace{0.5cm}

{\small \it  Key Laboratory for Quantum Information and Measurements
of Education Ministry,  Department of Physics, Tsinghua University,
Beijing 100084, P. R. China.}
\end{center}

A scheme of left-handed metamaterial (LHM) composed of
superconducting quantum interference devices (SQUIDs) and conducting
wires is proposed. The permeability of a probe field can be smoothly
tuned over a wide range with another electromagnetic (coupling)
field due to quantum interference effect. Similar to
electromagnetically induced transparency (EIT) of atomic systems,
the absorption of the probe field  can be strongly suppressed even
in the case of negative permeability. There are two passbands of
negative refractive index with low loss, which can be tuned with the
coupling field.

\begin{flushleft}

\end{flushleft}

Recently, research on left-handed materials (LHMs)\cite{veselago}
has attracted considerable  attentions. The LHMs are a kind of
metameterial which have negative permittivity and negative
permeability , which lead to negative refractive index in a narrow
frequency band. Many novel  phenomena can occur on LHM such as
superprism, perfect flat lens,  inverse light pressure,  and reverse
Doppler and Vavilov-Cherenkov effects .  Large nonlinearity can also
occur, e.g. the bistable transition of permeability from positive to
negative\cite{classical}. The quantum phenomena in LHMs have also
attracted attentions such as the modified spontaneous emission of
atoms in LHM \cite{spon1, spon2}.  However,  the background of the
ordinary LHM such as array of conductor lines and slit-ring
resonators (SRRs) is a 'classical' system where the negative
$\epsilon$ and negative $\mu$ arise from the classical plasma
oscillations. In this sense, the ordinary LHMs are quite different
from quantum systems such as atomic gases. On the other hand, in
ordinary situations, the magnetic response of atomic gases to a
laser field is too weak to generate negative permeability. How to
enhance the magnetic response of a quantum system is essential to
realize LHMs at optical frequency band.  There may be other ways to
obtain negative permeability, e.g., recently, a scheme of
electromagnetically-induced left-handedness in atomic gases is
proposed\cite{leftatom}, where large density of atoms is necessary
and special restrictions on frequencies of driven fields and atomic
transitions are required.  In contrast,  an artificial quantum
system with high magnetic sensibility can has much more advantage
than atomic systems to realize negative permeability.

In this letter we propose a new kind  of LHM composed of
superconducting rings with Josephson junctions and conductor wires.
For simplicity, the wires are assumed to be normal conductors. It
should be note that quite different from our work, recently analysis
and design of superconducting transmission lines are presented by
Salehi et al \cite{salehi}, and more recently, Ricci et al have
experimentally researched on a metamaterial that employs
superconducting Nb metals and low-loss dielectric materials,  in
which  case negative effective index passband are seen between 50MHz
to 18GHz \cite{superleft}.  The focus of our work is on the negative
permeability which arises from quantum feature of the composite. In
contrast with the 'classical' SRR, in our model the split of the
ring is replaced by a Josephson junction, which is essential to the
quantum feature of the LHM.  We will show analytically that due to
the quantum interference effects\cite{superEIT} the permeability can
be tuned over a wide range with an external microwave field. Also,
we will show that the absorption of the medium for the probe field
can be strongly suppressed even in the case of negative refractive
index.

 A scheme for supperconducting LHM
is show in Fig.1, where the composite is composed of superconducting
rings with Josephson junctions, and, an effective electric medium
with effective electric permittivity $\epsilon$ as a background,
which, e.g., can be an array of normal conducting wires. The radius
of the ring is denoted by $a$ , and the period of the array is
denotd by $d$. Schematic of the potential energy and the first six
eigen energies of the SQUID is shown in Fig. 2. We assume a probe
microwave field is interacting with the composite. The wavelength of
the field is assumed to be much longer than the period, i.e.,
$d<<\lambda$. For a classical cylindrical SRR system, the
permeability $\mu$ can be given by  the relation (according to
Ref.\cite{classical})
\begin{equation} {\bf B(\omega)}={\bf H}_x(\omega)+FH'(\omega)
,\end{equation}
 where $H_x(\omega)$ is the alternating external magnetic field and $H'(\omega)$
  is the additional magnetic field induced by $H_x(\omega)$, which determines the
   magnetization of the
composite, and $F=\pi a^2/d^2$ is the fraction of the structure with
$a$ being the radius of the ring and $d$ being the periodicity of
the array. Also, $a<<d<<\lambda$ is assumed.

 The permeability at angular frequency $\omega$ can be
given by
  \begin{equation} \mu(\omega)=1+F\f{\phi(\omega)}{\phi_x(\omega)}
\label{muphi},\end{equation} where $\phi(\omega)$ is the flux of
frequency $\omega$ induced by the external microwave field, i.e.
$\phi(\omega)=H's$ with $s=\pi a^2$ being the area of the ring. For
the quantum LHM here, the SRRs are replaced by the SQUIDs, then
$\phi (\omega)$ becomes an operator although the external driven
fields are assumed to be classical. In order to calculate $\mu$,
$\phi (\omega) $ in Eq. (\ref{muphi}) should be replaced by the
quantum averaging of it, i.e. $<\phi(\omega)>$.

The Hamiltonian of a SQUID (a ring with a Josephson junction) can be
given by\cite{zhou}
\begin{equation}
 H_0=-\frac{\hbar}{2m}\frac{\partial}{\partial x^2}+V(x)
\end{equation}
with the potential of the SQUID being

\begin{equation}
V(x)=\f{1}{2}m\omega_{LC}^2(x-x^\prime)-\f{1}{4\pi^2}m\omega_{LC}^2\beta
 cos(2\pi x)
\end{equation}
where $x=\phi/\phi_0$, $m=C\phi_0^2$, $\omega_{LC}^2=\f{1}{LC}$,
$\beta=2\pi LI_c/\phi_0$, and $x'=\phi_x/\phi_0$. Here $\phi$ is the
total flux in the ring, $L$ is the ring inductance, $\phi_x$ is an
external applied magnetic flux to the SQUID, $I_c$ is the critical
current of the junction, $C$ is the capacitance of the junction, and
$\phi_0(=h/2e)$ is the flux quantum. We consider a realistic SQUID
system which can be described by use of the parameters as in the
work of Zhou etal \cite{zhou}, where $L=100pH$, $C=40fF$, and
$I_c=3.95\mu A$, leading to $\omega_{LC}=5\times10^{11}rad/s$, and
$\beta=1.2$. The external DC magnetic field parameter $x'$ is taken
to be $-0.501$.

The interaction between the SQUID and microwave fields, which are
assumed to be linearly polarized with their magnetic field
perpendicular to the plane of the SQUID ring, is described by the
time dependent potential
\begin{equation}
 V_{int}(x,t)=m\omega_{LC}^2(x-x')(\varepsilon cos\omega t +\varepsilon_c cos\omega_c
 t)
.\end{equation}
In the interaction picture ,the dynamics of the
system is governed by the Schr{\"o}dinger equation
\begin{equation}
 i\hbar\f{\partial}{\partial t}|\psi(t)>=V_{int}|\psi(t)>
\label{s}.\end{equation}
 Here $\varepsilon$, $\varepsilon_c$ are microwave magnetic
flux of the probe field and the coupling field in units of $\phi_0$.
The frequency of the probe field and the coupling field are chosen
near resonant with transition $|0>-|n>$ and $|1>-|n>)$ respectively,
where $|n>$ denotes the $n$-th  eigen state of $H_0$. This is
usually referred to as the three-level approximation. In this case,
the wave function can be written as
\begin{equation}
 |\psi(t)>=c_0(t)|0>+c_1(t)|1>+c_n(t)|n>
,\end{equation} and the interaction Hamiltonian after the
rotating-wave approximation, can be written as \cite{zhou}
\begin{equation}
 V_{int}=\hbar(\Omega e^{i\delta t}|0><n|+\Omega_c e^{-i\delta_c t})|1><n|+
 h.c.
-i\f{\gamma_1}{2}|1><1|-i\f{\gamma_n}{2}|2><2|).
\end{equation}
Here the Rabi frequency $\Omega$ and $\Omega_c$ are defined as
\begin{equation}\Omega=-x_{0n}m\omega_{LC}^2\epsilon/\hbar, \, \, \, \, \, \Omega_c=-x_{1n}m\omega_{LC}^2
\epsilon_c/\hbar ,\end{equation}
 where $x_{0n}$  and $x_{1n}$ are
defined as $<0|x|n>$ and $<1|x|n>$ respectively. Also, here
$\delta=\omega-\omega_{0n}$, $\delta_c=\omega_c-\omega_{1n}$,
$\gamma_1$ and $\gamma_n$ are the background decay rate of the state
$|1>$ and $|n>$. The equations for amplitudes $c_0$, $c_1$, and
$c_n$ can be easily obtained from Eq.(\ref{s}) as
\begin{equation}
\begin{array}{lcr}
i \f{dc_0}{dt}=\f{\Omega}{2}c_n\\
i \f{dc_n}{dt}=-(\delta+i\f{\gamma_n}{2})c_n+\f{\Omega}{2}c_0+\f{\Omega_c}{2}c_1\\
i\f{dc_1}{dt}=-(\delta-\delta_c+i\f{\gamma_1}{2})c_1+\f{\Omega_c}{2}c_n
\end{array} ,
\label{amplitude}
\end{equation}
We assume that $\gamma_1<<\gamma_n$, where $\gamma_n$ can be taken
to be about 1GHz according to Ref. \cite{decay,superEIT}. We assume
that initially the system is prepared in the ground state $|0>$, and
for simplicity, we assume the probe field is weak enough, i.e.
$\Omega<<\gamma_n$,$\Omega<<\Omega_c$. In this case the steady-state
solution can be obtained as in ordinary EIT
systems.

It is easy to prove that the mean value of the flux
($<\phi(\omega)>$) can be given by
\begin{equation} \f{<\phi(\omega)>}{\phi_x(\omega)}=-\alpha \f{c_0^*c_1}{\Omega}
, \label{phiomega}
\end{equation}
 where $\alpha=\f{m\omega_{LC}^2
x_{0n}^2}{\hbar}=\f{\phi_0^2}{\hbar L} x_{0n}^2$. Here $\alpha$ is a
parameter measures the sensibility of the effective medium to the
probe field, which has the dimensions of frequency (Hz). For $n=4$,
$x_{04}=5.39798\times10^{-3}$, which is given by
Ref.\cite{paspalakis, Zafiris}.  Taking $L=100pH$, we obtain
$\alpha=11.815$GHz. This value will be used in the following
analysis.

We first consider the case of $\Omega_c=0$, i.e. of  two-level
system (non-EIT system). According to
Eq.(\ref{muphi}),(\ref{phiomega}) and (\ref{amplitude}), the
permeability can be give by
\begin{equation}
\mu=1-\f{F}{2}\f{\alpha}{\delta+i\gamma_4/2} \label{munon}
\end{equation},
It is clearly a Lorentz profile. When the probe field is near
resonant with the transition $|0>-|n>$, strong absorption will
occur. The absorption becomes sharp with the decrease of $\gamma_4$.
Similar to classical SRR composite, negative $\mu$ can occur (see
Fig. 3(a)) if $\alpha$ is large enough. The condition for the
negative $\mu$ can be easily deduced from Eq. \ref{munon} as that
$F\alpha>2\gamma_4$. Also,  it is clear that there is only one
frequency band of negative Re$(\mu)$.  However , it is worth to note
that the $\mu$-spectrum for the quantum composite here is similar
but not the same as that in classical composite\cite{classical}.

If the coupling field is applied to the system ($\Omega_c \not=0$),
the magnetic response could be significantly modified. For a strong
coupling field, i.e. $\Omega_c>>\Omega$, $\gamma_4$, from
Eq.(\ref{muphi}),(\ref{phiomega}) and (\ref{amplitude}), the
permeability can be given by
\begin{equation}
\mu=1-\f{F}{2}\f{\alpha
\(\delta-\delta_c+i\f{\gamma_1}{2}\)}{\(\delta-\delta_c+i\f{\gamma_1}{2}\)
\(\delta+i\f{\gamma_4}{2}\)-\f{\Omega_c^2}{4}}. \label{muEIT}
\end{equation}

The spectrum of $\mu$ in three-level case is show in Fig. 3(b). It
can be easily seen that the transparency occurs when
$\delta=\delta_c=0$, which is similar to EIT of atomic system except
that here the coupling field is used to control the permeability
$\mu$ instead of the permittivity $\epsilon$ of the probe field.
Superconductive analog to electromagnetically induced transparency
that utilizes superconductive quantum circuit designs of present day
experimental consideration has also been investigated by Muralial et
al \cite{superEIT}. The width of the transparency window can be
defined by the frequency distance between the two absorption peaks
which is about $\Omega_c$ due to Autler-Townes splitting. The
frequency band for low absorption and large refraction can occur
over a wide range for large $\Omega_c$. On the contrary, the group
velocity of the probe field can be reduced to near zero as in atomic
EIT systems. If $F\alpha$ is very small, $\mu$ is positive for all
probe detunings ($\delta$), which can be easily seen from
Eq.(\ref{muEIT}). If $F\alpha$ is large enough, however, negative
Re($\mu$) can occur near the transparency window. The condition for
negative $\mu$ can be deduced from Eq. (\ref{muEIT}) as that
\begin{equation}
F\alpha>2\gamma_4.
\end{equation}
This condition is the same as that for nonEIT case, but the
frequency band for negative $\mu$ in the EIT case and nonEIT case is
quite different. In the EIT case there are two frequency bands for
negative $\mu$, which can be given by
\begin{equation}
f_-(x_-)<\delta/\gamma_4<f_-(x_+);\, \, \, \, \, \, \, \,
f_+(x_-)<\delta/\gamma_4<f_+(x_+),
\end{equation}
where
\begin{equation}
f_{\pm}(x)=\f{x\pm \sqrt{x^2+\Omega_c^2/\gamma_4^2}}{2},
\end{equation}
and
\begin{equation}
x_\pm=\f{1}{2}(g\pm\sqrt{g^2-1})
\end{equation}
with $g=\f{F\alpha}{2\gamma_4}(>1)$

For example, we consider the case of $\alpha=11.815$GHz as above. If
taking $F=\pi\times0.14$, we can obtain that $F\alpha/2=2.5982$GHz.
In this case negative Re$(\mu)$ can occur if the condition
$\gamma_4<2.5982$GHz is met.

It should be note that, when $\delta=\delta_c$ and $\gamma_1=0$,
exactly zero absorption occcur (Im$(\mu)=0$), but in this case
Re$(\mu)=+1$. When Re$(\mu)$ is negative, Im$(\mu)$ is always
nonzero, i.e. negative refractive index without loss can never
occur. However, the absorption can be strongly suppressed near the
condition of $\delta=\delta_c$. It is similar to ordinary EIT in
atomic systems, but here the zero absorption occurs due to magnetic
response instead of electric response, and, here, the magnetic is so
strong that negative Re$(\mu)$ can occur near EIT condition. The
permeability ($\mu$) can be smoothly tuned over a wide range with
the coupling detuning $\delta_c$ (see Fig. 4) and its Rabi frequency
$\Omega_c$ (see Fig.(5)). It is worth to note that the tunability is
due to the quantum interference effect which is absent in classical
normal conductor systems.  The permeability $\mu$ and refractive
index $n$ can be smoothly tuned by the coupling field over a wide
range, and the continuous transition from Re$(\mu)=1$ to
Re$(\mu)=-1$ is possible. From the $\mu$-spectrum it is easy to see
that there are four frequencies for Re$(\mu)=-1$, but only two of
them are for low loss (small Im$(\mu)$), one of which is in EIT
window, while the other one is out of EIT window. It is should be
note that both of them can be tuned with the coulpling field.  For
this two frequencies, Im$(\mu)\rightarrow 0$ when $\gamma_4
\rightarrow 0$ and $\gamma_1 \rightarrow 0$.

In order to calculate the refractive index, the permittivity
$\epsilon$ and permeability $\mu$ can be written as
\begin{equation}
\epsilon=|\epsilon|e^{i\phi_e},  \mu=|\mu|e^{i\phi_m}
,\end{equation} where $0<\phi_e<\pi$,  $0<\phi_m<\pi$. Then the
refractive index $n$ can be given by \cite{spon1}
\begin{equation}
n=\sqrt{|\epsilon||\mu|}e^{i(\phi_e+\phi_m)/2}
\end{equation}
For the sake of simplicity, we consider the case where the effective
electric permittivity $\epsilon$ is generated by an array of normal
conducting classical current wires\cite{spon1,classical}. In this
case
\begin{equation}
\epsilon=1+\f{\omega_{pe}^2}{\omega_{Te}^2-\omega^2-i\omega\gamma_e}
\label{epsilon}
\end{equation}
where $\omega_{pe}$ and $\omega_{Te}$ are the parameters which
measure the effective plasma oscillations and $\gamma_e$ measures
the loss of the wires. The refractive index spectrum of the
composite is shown in Fig. (6), where (a) is for $\epsilon(\delta)$,
(b) is for $\mu(\delta)$, and (c) is for $n(\delta)$. For
simplicity, we consider the frequency band of negative $\epsilon$
which is far from resonance, in which case $\epsilon$ is slowly
varing with $\delta$, and Im$(\epsilon)$ is small. We find in this
case there are two minima on the Im$(n)$-spectrum. The absorption
minima arise from two reasons: (1) small Im$(\epsilon)$ and small
Im$(\mu)$; (2)Re$(\mu)$ and Re$(\epsilon)$ are simultaneously
negative. On the contrary, if Re$(\mu)Re(\epsilon)<0$, Im$(n)$ will
be large even if both of Im$(\mu)$ and Im$(\epsilon)$ are infinitely
small.   Two passbands of negative refractive index appear around
the two minima, both of them are sensitive to the coupling field.

In conclusion, we have  analytically investigated the metamaterial
composed of superconducting rings with Josephson junctions and
conducting wires. It is found that negative permeability for a probe
microwave field can occur when $g=\f{F\alpha}{2\gamma_4}>1$. There
are two passbands of negative refractive index with low loss, which
can be tuned with the coupling field.  To our knowledge, it is for
the first time that the quantum left-handed metamaterial being
composed of SQUIDs has been investigated. One of the advantages of
this kind of LHM is that the negative refractive index with low loss
is easy to obtain due to that the permeability can be smoothly tuned
over a wide range including $\mu=\pm 1+i\varepsilon$
($\varepsilon\rightarrow 0+$). In the quantum composite, some new
physics could be found, such as that associated with transient
properties of the left-handedness, large nonlinearity due to the
quantum interference effect and that due to strong magnetic response
of the composite. Also, the tunability of the composite via quantum
interference effect could be used in quantum information process
\cite{information}. On the other hand, the LHM with SQUIDs can be
regarded as a bridge between the 'classical' LHMs (e.g. composed of
SRRs and wires) and the 'quantum' systems composed of microscopic
particles(e.g. atoms or quantum dots). Hence, further research on
the quantum LHM could facilitate the realizing of the tunable LHMs
of high-frequency (towarding to optical) band.

\section*{ACKNOWLEDGMENT}
This work is supported by the National Nature Science Foundation of
China (Grant No. 10504016) and funded by the State Key Development
Program for Basic Research of China(Grant No. 001CB309308).
\newpage



\newpage
\begin{center}
{\bf Figure Captions }
\end {center}
\begin{flushleft}
Fig. 1.  Schematic of the composite metamaterial structure composed
of superconducting rings with Josephson junctions (SQUIDs) and the
effect electric medium such as conductor wires. Each SQUID is an
artificial $\Lambda$-configuration three-level system which coupled
with a probe microwave field and a coupling microwave field.\\[0.4cm]

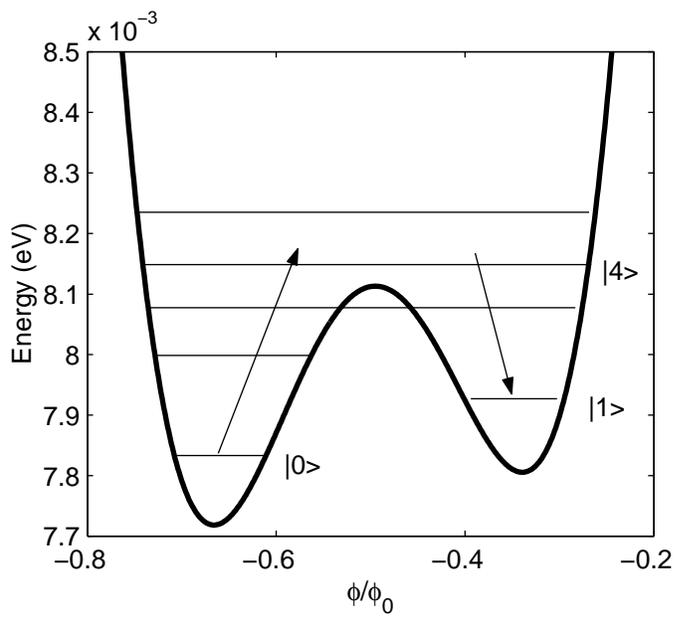
Fig. 2.   Schematic of the potential energy and the first six eigen
energies of the SQUID. The energies of the ground state $|0>$,
meta-stable state $|1>$, and the excited state $|4>$
are 7.81984mev, 7.90183mev, and 8.14057mev, respectively.\\[0.4cm]

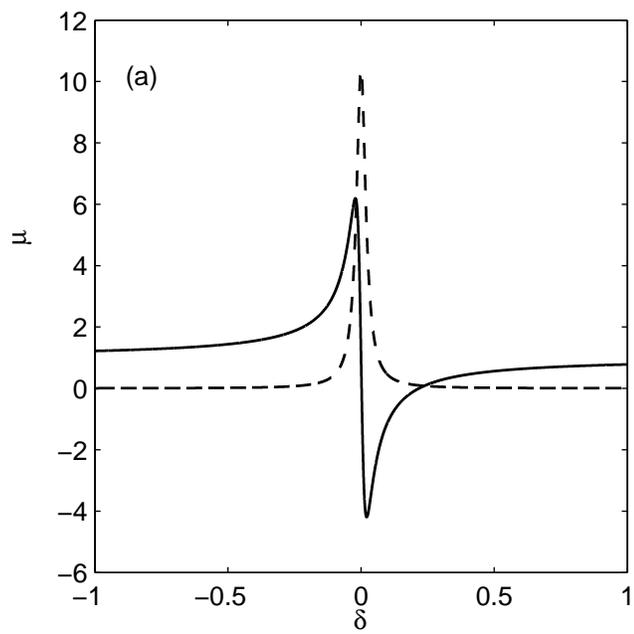
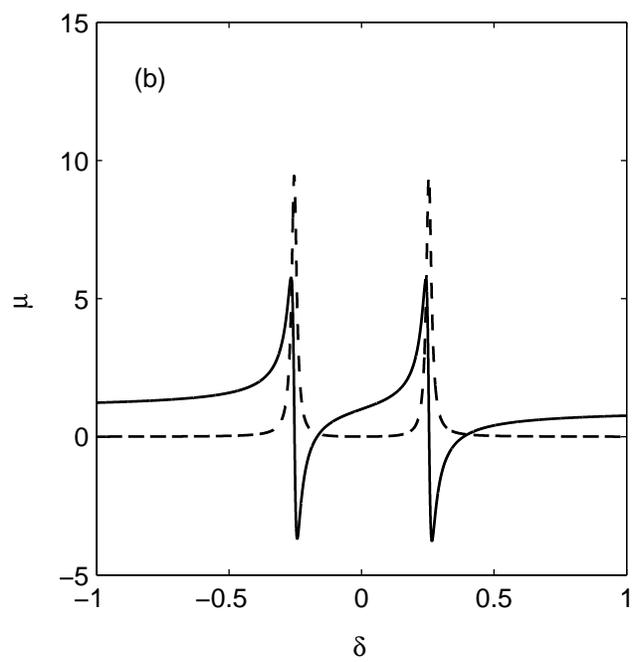
Fig. 3.  Real part (solid curve) and imaginary part (dashed curve)
of the permeability ($\mu$) versus the probe detuning $\delta$ for
$\delta_c=0$, where $F=\pi\times 0.14$, $\gamma_4=0.0423$,
$\gamma_1=0.1\gamma_4$, $\beta=1.2$, $x'=-0.501$. (a) is for
$\Omega_c=0$ and (b) is for $\Omega_c=0.5078$. All parameters are in
units of $\alpha$, which is taken to be 11.815GHz.  \\[0.4cm]

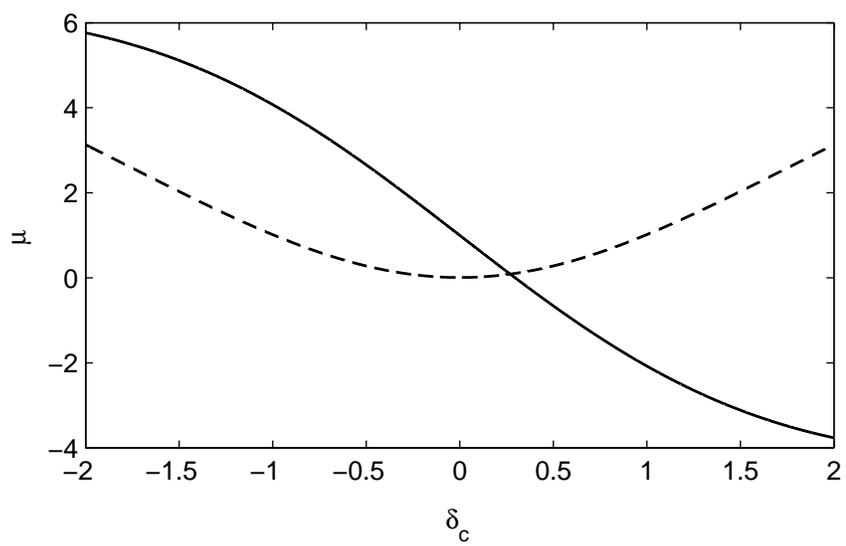
Fig. 4.  Real part (solid curve) and imaginary part (dashed curve)
of the permeability ($\mu$) versus the coupling detuning $\delta_c$
for
the case of $\delta=0$. The parameters are the same as that in Fig. 3\\[0.4cm]

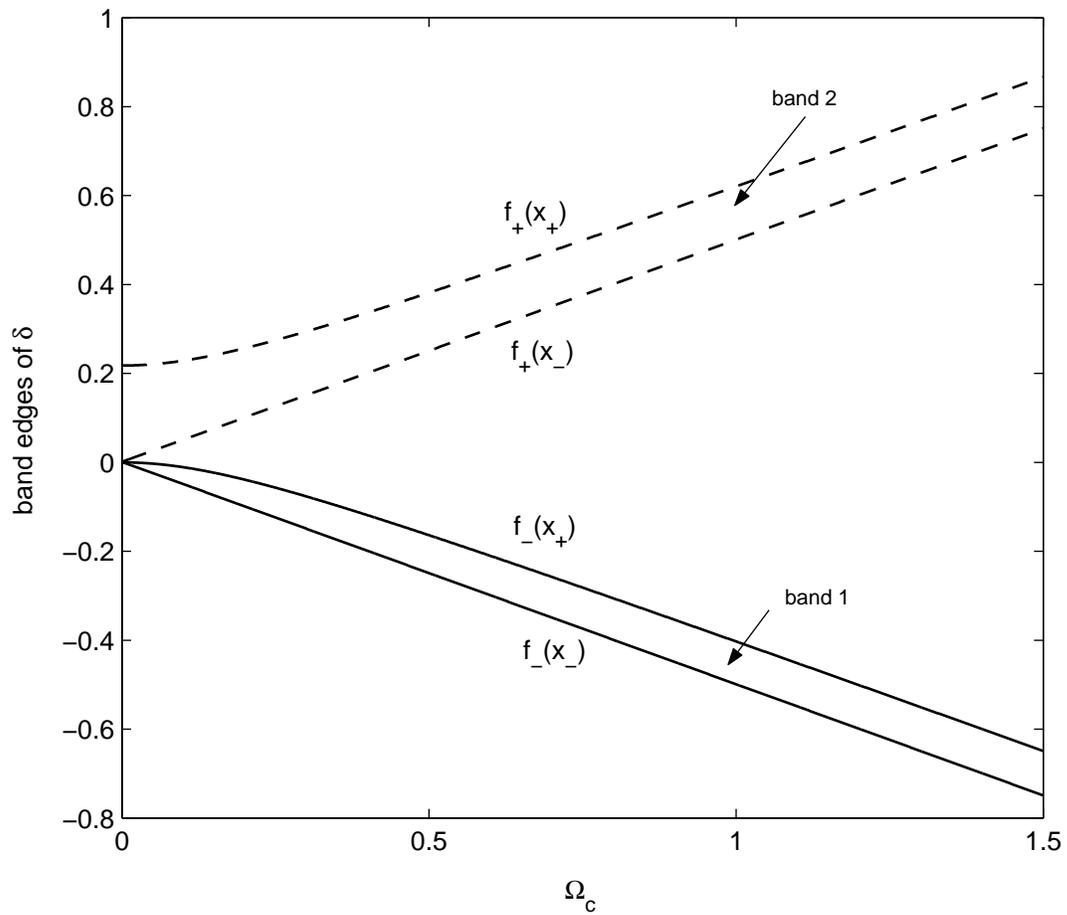
Fig.5.  The band edges of negative Re$(\mu)$ versus the Rabi
frequency of the coupling field $\Omega_c$. Negative Re$(\mu)$
occurs when $\delta$ is between the two solid curves (band1) or that
between the
dashed curves(band2). The parameters are the same as that in Fig.3.\\[0.4cm]

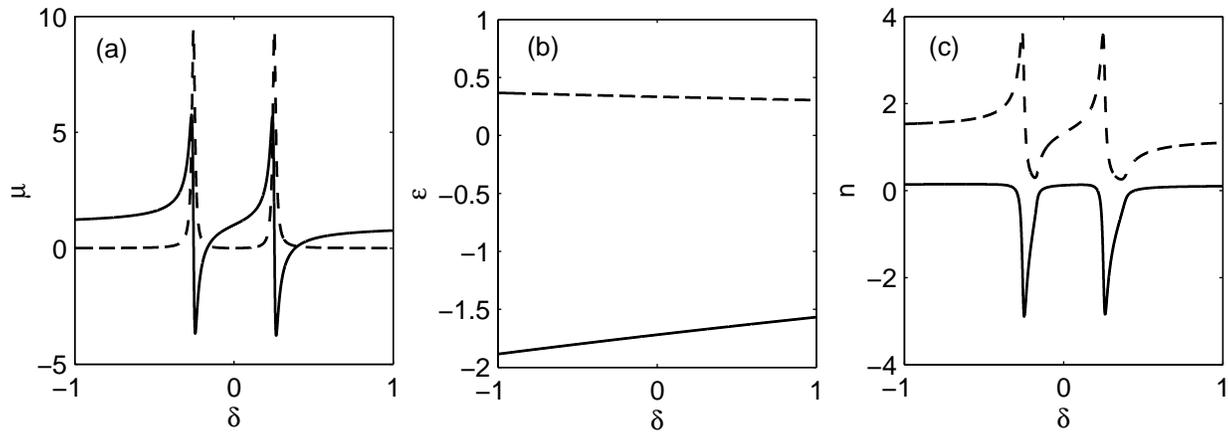
Fig.6.  The spectrum of permittivity(a), permeability(b), and
refractive index(c), where $\omega_{pe}=1.5\omega_{04}$,
$\omega_{Te}=0.43\omega_{04}$, $\gamma_e=0.1\omega_{04}$. The solid
line is for the real part of them, while the dashed line is for the
imaginary part of them. Here
$\omega_{04}=48.727$GHz. Other parameters are the same as that in Fig.3.\\[0.4cm]

\end{flushleft}

\begin{thebibliography}{99}

\bibitem {veselago} V. G. Veselago, Usp. Fiz Nauk {\bf 8}, 2854 (1967)
[Sov. Phys. Usp. {\bf 10}, 509 (1968)]

\bibitem {classical}  A. A. Zharov, I. V. Shadrivov, and Y. S. Kivshar, Phys. Rev. Lett.,
{\bf 91} 037401  (2003)

\bibitem {spon1} H. T. Dung , S. Y. Buhmann, L. Knoll et al., Phys. Rev A {\bf 68}, 043816 (2003)

\bibitem {spon2} J. Kastel and M. Fleischhauer, Phys.Rev. A {bf 71} 011804 (2005)

\bibitem {leftatom}  M. {\"O}. Oktel and {\"O}. E. M{\"u}stecaplioglu, Phys. Rev. A {\bf 70},
053806 (2004).

\bibitem {salehi} H. Salehi, A. H. Majedi, R. R. Mansour,
IEEE TRANSACTIONS ON APPLIED SUPERCONDUCTIVITY {\bf 15},996 (2005)


\bibitem {superleft} M. Ricci, N. Orloff, and S. M. Anlage,
Appl. Phys. Lett., {\bf 87} 034102 (2005)

\bibitem{superEIT} K.V. R.M. Murali, Z. Dutton, W. D. Oliver, D. S. Crankshaw, and T. P.
Orlando, Phys .Rev. Lett. {\bf 93} 087003 (2004)

\bibitem {zhou} Z. Zhou, S. I. Chu, and S. Han, Phys. Rev. B{\bf 66}, 054527 (2002)

\bibitem {decay}Y. N. Ovchinnikov, P. Silvestrini, V. Corato, and S. Rombetto, Phys. Rev. B
{\bf 71}, 024529  (2005).

\bibitem {paspalakis} E. Paspalakis and N. J. Kylstra, J. Mod. Opt. {\bf 51}, 1679 (2004);


\bibitem {Zafiris} N. Aravantinos-Zafiris and E. Paspalakis, Phys. Rev. A {\bf 72} 014303 (2005).


\bibitem {information} C. P. Yang, S. I. Chu, and S. Han, Phys. Rev. Lett. {\bf 92} 117902 (2004).





\end{thebibliography}
\end{document}